\documentclass[a4paper,fleqn]{mnras}

\usepackage[T1]{fontenc}
\usepackage{ae,aecompl}

\usepackage{graphicx}	% Including figure files
\usepackage{amsmath}	% Advanced maths commands
\usepackage{amssymb}	% Extra maths symbols
\usepackage{subfig}     % Including subfigures
\title[Proper motions of 15 pulsars]{Proper motions of 15 pulsars: a comparison between Bayesian and frequentist algorithms}

\author[L. Li et al.]{
L. Li,$^{1,2,3}$\thanks{E-mail: lilin@xao.ac.cn}
N. Wang,$^{1,4}$
J. P. Yuan,$^{1,4}$
J. B. Wang,$^{1,4}$
G. Hobbs,$^{5}$
L. Lentati,$^{6}$
\newauthor and R. N. Manchester$^{5}$
\\
$^{1}$Xinjiang Astronomical Observatory, Chinese Academy of Sciences, 150 Science 1-street, Urumqi, Xinjiang 830011, China \\
$^{2}$University of Chinese Academy of Sciences, 19A Yuquan Road, Beijing 100049, China \\
$^{3}$School of Physics Science and Technology, Xinjiang University, Urumqi, Xinjiang 830046, China \\
$^{4}$Key Laboratory of Radio Astronomy, Chinese Academy of Science, Nanjing 210008, China \\
$^{5}$CSIRO Astronomy and Space Science, Australia Telescope National Facility, PO Box 76, Epping, NSW 1710, Australia \\
$^{6}$Astrophysics Group, Cavendish Laboratory, JJ Thomson Avenue, Cambridge, CB3 0HE, UK \\
}

\date{Accepted ... Received ...; in original form 2016 March }

\pubyear{2016}

\begin{document}
\label{firstpage}
\pagerange{\pageref{firstpage}--\pageref{lastpage}}
\maketitle

\begin{abstract}

We present proper motions for 15 pulsars which are observed regularly by the Nanshan 25-m radio telescope. Two methods, the frequentist method (Coles et al. 2011) and the Bayesian (Lentati et al. 2014) method, are used and the results are compared. We demonstrate that the two methods can be applied to young pulsar data sets that exhibit large amounts of timing noise with steep spectral exponents and give consistent results. The measured positions also agree with very-long-baseline interferometric positions. Proper motions for four pulsars are obtained for the first time, and improved values are obtained for five pulsars.

\end{abstract}


%%%%%%%%%%%%%%%%%%%%%%%%%%%%%%%%%%%%%%%%%%%%%%%%%%

%%%%%%%%%%%%%%%%% BODY OF PAPER %%%%%%%%%%%%%%%%%%

\section{Introduction}

Pulsar velocities are usually determined by measuring their proper motions and distances. Lyne \& Lorimer (1994) and Hobbs et al. (2005) analysed a large number of pulsar proper motions and demonstrated that their transverse velocities are typically several hundred km$\,$s$^{-1}$. Various explanations have been proposed for these high velocities, including a postnatal electromagnetic rocket mechanism (Harrison \& Tademaru 1975), asymmetric neutrino emission in the presence of super-strong magnetic fields (Lai \& Qian 1998), hydrodynamical instabilities in the collapsed supernova core (Lai \& Goldreich 2000) and the asymmetric explosion of $\gamma$-ray bursts (Cui et al. 2007). Knowledge of pulsar proper motions and velocities is essential for many aspects of pulsar and neutron star astrophysics including determinations of the birth rate of pulsars, associations with supernova remnants and the Galactic distribution of the progenitor population.   Unfortunately only $\sim10\%$ of the known pulsars currently have measured proper motions (Manchester et al. 2005)\footnote{http://www.atnf.csiro.au/research/pulsar/psrcat/, the Australia Telescope National Facility (ATNF) Pulsar Catalogue}. It is therefore highly desirable to determine more proper motions and to confirm the accuracy of previous publications.

Determining proper motions is not trivial. The most common methods\footnote{Other methods to determine the proper motion include optical techniques (Trimble 1971; Wyckoff \& Murray 1977; \"Ogelman et al. 1989; Mignani et al. 2000) and using the scintillation properties of pulsars(e.g., Wang et al. 2005).} are 1) using very-long-baseline interferometry (e.g., Lyne et al. 1982; Harrison et al. 1993; Brisken et al. 2003; Chatterjee et al. 2009; Yan et al. 2013), but this is usually only possible for relatively close-by and bright pulsars or 2) using the pulsar timing method. The use of the pulsar timing method for determining proper motions was first presented by Manchester et al. (1974) who detected the proper motion for PSR~B1133+16 using a 4-year timing data set. Recent proper motions have been published by numerous groups including Nice \& Taylor (1995), Hobbs et al. (2004), Zou et al. (2005), Gonzalez et al. (2011), Desvignes et al. (2016), Matthews et al. (2016).

As data sets increased it became clear that unmodelled irregularities in the pulsar timing residuals (pulsar timing noise) could bias the reported proper motions. This was commonly dealt with simply by increasing the uncertainties reported by the pulsar timing software or by using high-order pulse-frequency derivatives (e.g., Zou et al. 2005) or harmonic whitening (Hobbs et al. 2005) to absorb the noise. Coles et al. (2011) showed that both of these methods still led to biased parameters and incorrect uncertainty estimates. They presented a new method, based on a generalised-least-squares fitting procedure, in which the spectrum of timing noise was estimated using an analytic model. The temporal correlations in the timing residuals were then subsequently accounted for in the timing model fitting. This method has now been used to determine the proper motion of millisecond pulsars (Reardon et al. 2016). Lentati (2014) demonstrated a similar method in which the noise properties of the data and the pulsar parameters were simultaneously determined using a Bayesian methodology (Desvignes et al. 2016). Recently Babak (2016) described the comparison of noise models from Bayesian and frequentist methods and got the consistent upper limits on continuous gravitational waves depended on the two algorithms. Neither of these methods have been applied to a large sample of young pulsars (they were both developed to study millisecond pulsars). In this paper we demonstrate that these two methods can be applied to noisy young pulsar data sets and we compare the results obtained from the methods.

\section{Observation and data reduction}

In this paper we describe the analysis of timing observations of 15 pulsars that were obtained with the Nanshan 25-m radio telescope of the Xinjiang Astronomical Observatory (XAO) as part of their regular timing program. The details of observing system have been described by Wang et al. (2001).  In brief, observations are carried out in the 18\,cm band.  The backend system consists of an analogue filterbank (AFB) system and, since 2010, simultaneous observations with a digital filterbank system (DFB).   The pulsars described in this paper were observed from 2000 January to 2013 Dec (apart from a few for which observations started in July, 2002). They are all typical, isolated pulsars with characteristic age of $10^{5}\sim10^{7}\,$yr and within a distance of 7$\,$kpc. Most have been selected because they are relatively bright in the 18\,cm observing band.  The pulsars are each observed approximately 2 to 4 times per month.  We list the basic parameters of these pulsars and the details of our data set in Table~\ref{tab:pama}.

\begin{figure}
\centering
 \includegraphics[angle=0, width=0.38\textwidth]{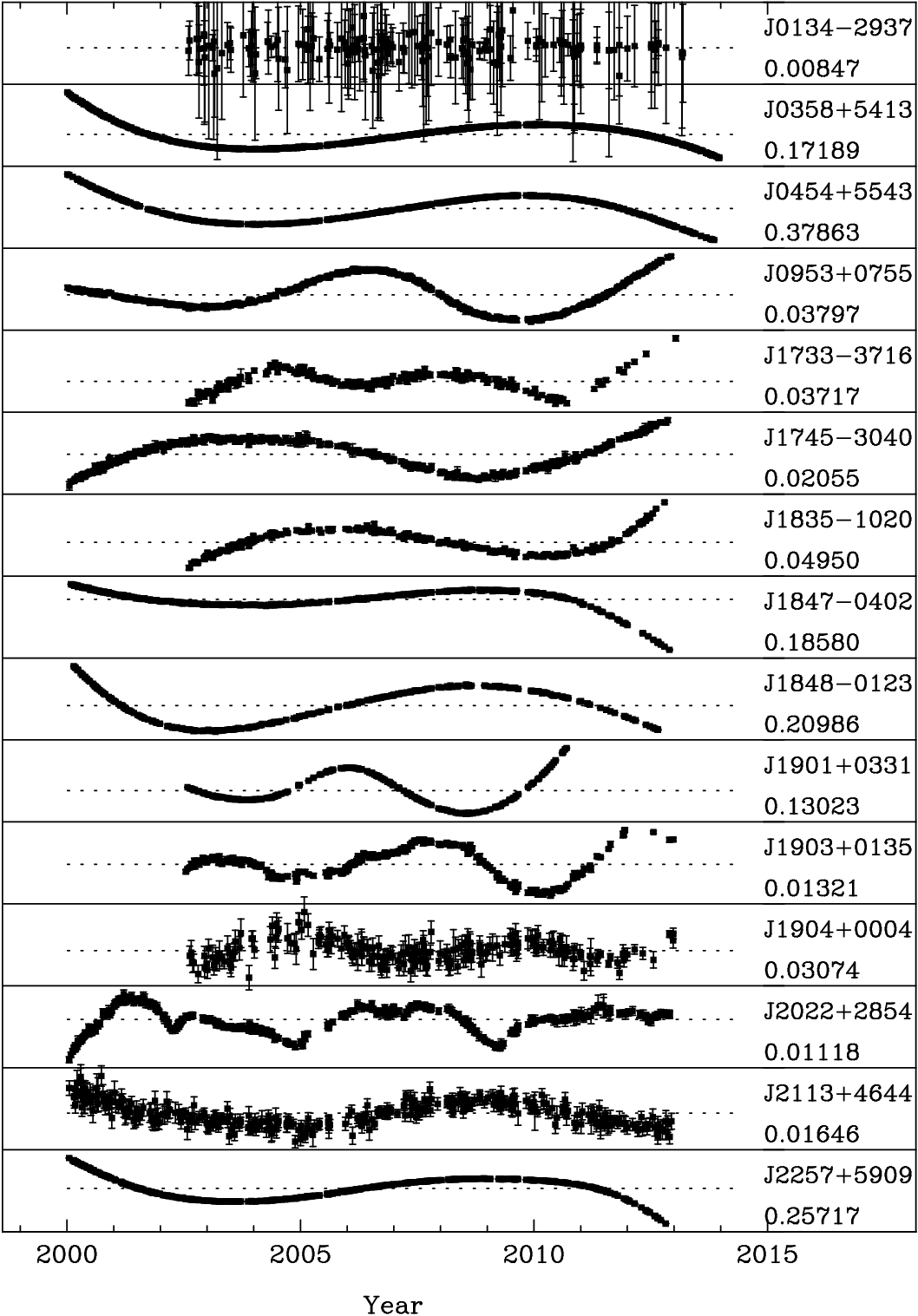}
\caption{ Timing residuals of 15 pulsars from 2000 to 2013. They arise from the frequentist analysis after fitting for the pulse frequency and its first derivative with the fixed astrometric parameters. The plot y-range in seconds is given under the pulsar name.}
\label{fig:res}
\end{figure}

Each pulsar is observed for $\sim$4--16 minutes and the resulting data are recorded to disk.  The \textsc{psrchive} package (Hotan et al. 2004) is used for off-line data reduction with an initial set of parameters being obtained from the ATNF pulsar catalogue (Manchester et al. 2005). After dedispersing and removing radio-frequency-interference (RFI), the data are summed in frequency, time and polarisation to produce a mean pulse profile.  An analytic template is obtained by fitting one or more Gaussian profiles to the mean pulse profile.  The template is then cross-correlated with each observation to obtain pulse times-of-arrival (ToAs). The \textsc{tempo2} (Hobbs et al. 2006, Edwards et al. 2006) timing package is then used to convert the local ToAs to the arrival times at the solar system barycentre with the Jet Propulsion Laboratory (JPL) planetary ephemeris DE421 (Folkner et al. 2008). Our observations are referred to Terrestrial Time as realised by BIPM2013\footnote{http://www.bipm.org}.  We use the standard procedures within the \textsc{tempo2} software package to compute the timing residuals and improve the timing models using the weighted, least-square fitting procedure. As part of this process we account for the offset of the two different backend systems (AFB \& DFB). For simultaneous observations with both systems, the ToA with the smaller uncertainty is chosen for subsequent processing.  An integral MJD near the centre of the data span is chosen as the epoch of the position and pulse frequency.  For our preliminary analysis, we fit the pulsar rotational frequency and its first derivative.  This initial model is subsequently updated using the methods described below in order to determine (or estimate) the pulsar proper motions. Timing residuals for our sample after fitting for the pulse frequency and its first derivative are shown in Fig.~\ref{fig:res}, in which the measured astrometric parameters (positions and proper motions) are fixed.

\begin{table*}\scriptsize
\centering
\caption{Parameters for the 15 pulsars described in this paper. Each pulsar name is followed by the rotational parameters from the frequentist method. The remaining columns are, in turn, the dispersion measure and distance (NE2001) obtained from the ATNF pulsar catalogue, the number of ToAs, the epoch of the period determination, the data span in years and the MJD range, the mean ToA uncertainty and the weighted rms variation of the timing residuals.  }
 \label{tab:pama}
 \begin{tabular}{cclllccccccr}
 \hline\noalign{\smallskip}
PSR J & PSR B & $\nu$ & $\dot{\nu}$ & DM & Dist. & N$_{\rm {ToA}}$ & Epoch & Span & MJD range & $\sigma_{\rm {ToA}}$ & W$_{\rm {rms}}$  \\
      &       & (s$^{-1}$) & ($10^{-14}\,s^{-2}$) & (pc$\,$cm$^{-3}$) & (kpc) & & (MJD) & (yr) & & ($\mu$s) & (ms)   \\
 \hline\noalign{\smallskip}
J0134$-$2937 & -          & 7.3013160129300(7) & $-$0.4177838(14) & 21.806(6)  & 0.56 & 187 & 54428 & 10.6 & 52494-56361 & 115 & 0.3  \\
J0358$+$5413 & B0355$+$54 & 6.3945110925(3)    & $-$17.9709 (2)   & 57.1420(3) & 1.00 & 988 & 54096 & 14.0 & 51547-56645 & 93  & 3.3  \\
J0454$+$5543 & B0450$+$55 & 2.9348725506(12)   & $-$2.0479(9)     & 14.5943(9) & 1.18 & 458 & 54086 & 13.9 & 51547-56625 & 326 & 21.0 \\
J0953$+$0755 & B0950$+$08 & 3.95154897424(7)   & $-$0.35941(9)    & 2.96927(8) & 0.26 & 605 & 53908 & 12.9 & 51547-56269 & 65  & 2.4  \\
J1733$-$3716 & B1730$-$37 & 2.9621405983(1)    & $-$13.20051(16)  & 153.5(3)   & 2.78 & 218 & 54433 & 10.6 & 52495-56372 & 481 & 2.1  \\
\\
J1745$-$3040 & B1742$-$30 & 2.72158867832(8)   & $-$7.90408(7)    & 88.373(4)  & 0.20 & 403 & 53895 & 12.8 & 51549-56241 & 126 & 1.6  \\
J1835$-$1020 & -          & 3.30633617194(16)  & $-$6.4702(2)     & 113.7(9)   & 2.30 & 251 & 54369 & 10.2 & 52497-56216 & 434 & 2.6  \\
J1847$-$0402 & B1844$-$04 & 1.6728219903(4)    & $-$14.4665(3)    & 141.979(5) & 3.26 & 366 & 53902 & 12.9 & 51550-56255 & 695 & 16.2 \\
J1848$-$0123 & B1845$-$01 & 1.5164524451(5)    & $-$1.2011(5)     & 159.531(8) & 4.40 & 270 & 53868 & 12.7 & 51549-56187 & 452 & 36.8 \\
J1901$+$0331 & B1859$+$03 & 1.5256623417(17)   & $-$1.738(2)      & 402.080(12)& 7.00 & 205 & 54336 & 10.1 & 52486-56186 & 352 & 27.7 \\
\\
J1903$+$0135 & B1900$+$01 & 1.37116686526(3)   &$-$0.75849(5)     & 245.167(6) & 3.30 & 278 & 54379 & 10.4 & 52473-56284 & 346 & 2.1  \\
J1904$+$0004 & -          & 7.16719146836(6)   & $-$0.60642(11)   & 233.61(4)  & 5.74 & 217 & 54377 & 10.3 & 52469-56285 & 530 & 0.7  \\
J2022$+$2854 & B2020$+$28 & 2.912031729147(11) & $-$1.605114(17)  & 24.632(1)  & 2.10 & 448 & 53901 & 12.9 & 51547-56254 & 112 & 0.7  \\
J2113$+$4644 & B2111$+$46 & 0.98552727895(3)   & $-$0.06858(3)    & 141.26(9)  & 4.00 & 349 & 53915 & 12.9 & 51559-56271 & 1267& 2.7  \\
J2257$+$5909 & B2255$+$58 & 2.7155585938(9)    & $-$4.2426(8)     & 151.082(6) & 3.00 & 434 & 53908 & 12.9 & 51560-56257 & 339 & 14.7 \\
 \noalign{\smallskip}\hline
 \end{tabular}
\end{table*}

\section{Methodology}

Two current methods exist for determining unbiassed proper motions in the presence of pulsar timing noise.  One is the frequentist method described by Coles et al. (2011) and the other is the Bayesian procedure presented by Lentati et al. (2014).  We have used both methods to estimate the red noise properties along with a determination of the proper motion for each pulsar.

The frequentist method is implemented as follows within the \textsc{tempo2} software package:

\begin{itemize}
\item We model any excess white noise scatter in the residuals using the \textsc{efacEquad} plugin (these are defined by parameters known as EFACs and EQUADs, see, e.g., Wang et al. 2015)\footnote{ The new ToA uncertainty, $e^\prime$ is related to the original uncertainty $e$ by ${e^\prime}^2 = {\rm EFAC}^2 \times (e^2 + {\rm EQUAD}^2)$. }.
\item Form timing residuals using the initial model of the pulsar parameters and use the \textsc{spectralModel} plugin package for \textsc{tempo2} to estimate an analytic model for the spectrum of the red noise component of the timing residuals. The red noise is parameterized using an amplitude $A$, a spectral exponent $\alpha$ and a corner frequency $f_c$ as follows:
\begin{equation}\label{eq:pf}
 P(f) = \frac{A}{[1+(f/f_{\rm c})^2)]^{\alpha/2}}.
\end{equation}
\item With the analytic red noise model we fit for the pulsar parameters (including the proper motion) using the global least-square-fitting procedure as described by Coles et al. (2011).
\item The entire process is iterated with these new model parameters and the analytic red noise model improved as necessary until the results converge.
\item We record the final improved parameters (red noise parameters, positions and proper motions) for later analysis.
\end{itemize}

 The Bayesian method is applied to the original set of ToAs and original pulsar parameters as follows:

\begin{itemize}
\item It is essential that we provide prior information on the range of proper motions.  For many of these pulsars, previous measurements of proper motions have already been obtained and so we usually have a reasonable first guess of the possible range of proper motions.  For our final processing we assume a flat prior for proper motions between $-500\,$mas/yr and $+500\,$mas/yr.
\item We are not (for this paper) interested in the pulse period and so we analytically marginalise over these parameters and the pulsar positions.
\item We include a noise model that parameterises excess white noise in the residuals\footnote{TNGLOBALEF \& TNGLOBALEQ are used, the definition is ${e^\prime}^2 = ({\rm TNGLOBALEF} \times e )^2 + {\rm TNGLOBALEQ}^2$. Note that these are different from the EFAC and EQUAD described earlier.}.
\item We then run the \textsc{temponest} plugin package to \textsc{tempo2} and determined the most likely noise parameters and proper motions.
\end{itemize}

\section{Results}

\subsection{Updated positions}

The frequentist method provided us with positions and proper motions.  The measured positions are given in Table~\ref{tab:pos}. The pulsar J2000 and B1950 names are listed in the first two columns of the table. The next three columns present the corresponding epoch of the position and the right ascension and declination obtained in equatorial coordinates from our data analysis respectively. The final four columns display the parameters from the literature including the epoch, the position and the reference. In this and succeeding tables, the figure in parentheses indicates 1$\sigma$ uncertainty in the last quoted digit. The superscripts `F' and `c' separately denote results from our measurements using the frequentist method and from the catalogue respectively. We note that the previously published positions for most of the 15 pulsars in our sample have not been updated for 10 to 20 years. We also emphasise that the measured positions (slightly) depend upon the choice of the JPL solar system ephemeris that was used (our results are based on JPL DE421, the earlier results are likely to have used DE200 or DE405). A comparison between the earlier positions and our measurements to obtain the proper motions is carried out in section~\ref{sec:pm}.

\begin{table*}\scriptsize
\centering
\caption{Pulsar positions of 15 pulsars obtained by applying the frequentist method to our data and also obtained from the literature. The first two columns list the J2000 and B1950 names, which are followed by the epoch of the position measurement, the right ascension and declination obtained in equatorial coordinates from our data analysis. The corresponding parameters from the ATNF Pulsar Catalogue and the references are in following four columns. The references in Table~\ref{tab:pos} and Table~\ref{tab:pm_comp} are all listed here: (1) Hobbs et al. 2004; (2) Chatterjee et al. 2004; (3) Chatterjee et al. 2009; (4) Johnston et al. 1995; (5) Zou et al. 2005; (6) Morris et al. 2002; (7) Brisken et al. 2002. The figure in parentheses indicates the 1$\sigma$ uncertainty in the unit of the last quoted digit and superscripts `F' and `c' separately denote results from the frequentist method and the catalogue in this and subsequent tables. }
 \label{tab:pos}
 \begin{tabular}{cccllcllc}
 \hline\noalign{\smallskip}
PSR J     & PSR B   & Epoch$^{\rm {F}}$ & RA$^{\rm {F}}$(\rm J2000) & Dec.$^{\rm {F}}$(\rm J2000) & Epoch$^{\rm {c}}$ & RA$^{\rm {c}}$(\rm J2000) & Dec.$^{\rm {c}}$(\rm J2000) & Ref. \\
 & & (MJD) & (h:m:s) & ($\degr:':''$) & (MJD) & (h:m:s) & ($\degr:':''$) & \\
 \hline\noalign{\smallskip}
J0134$-$2937 & -          & 54428 & 01:34:18.6824(3)  & $-$29:37:17.042(7)  & 50846 & 01:34:18.6690(15) & $-$29:37:16.91(3)   & 1 \\
J0358$+$5413 & B0355$+$54 & 54096 & 03:58:53.7238(2)  & $+$54:13:13.784(3)  & 51544 & 03:58:53.7165(3)  & $+$54:13:13.727(5)  & 2 \\
J0454$+$5543 & B0450$+$55 & 54086 & 04:54:07.7833(9)  & $+$55:43:41.345(13) & 52275 & 04:54:07.7506(1)  & $+$55:43:41.437(2)  & 3 \\
J0953$+$0755 & B0950$+$08 & 53908 & 09:53:09.3069(8)  & $+$07:55:36.37(3)   & 46375 & 09:53:09.3097(19) & $+$07:55:35.75(8)   & 1 \\
J1733$-$3716 & B1730$-$37 & 54433 & 17:33:26.7628(18) & $-$37:16:55.23(9)   & 48378 & 17:33:26.754(5)   & $-$37:16:54.6(3)    & 4 \\
\\
J1745$-$3040 & B1742$-$30 & 53895 & 17:45:56.3127(4)  & $-$30:40:23.19(4)   & 52384 & 17:45:56.305(2)   & $-$30:40:23.5(2)    & 5 \\
J1835$-$1020 & -          & 54369 & 18:35:57.5784(10) & $-$10:20:04.50(6)   & 51493 & 18:35:57.551(12)  & $-$10:20:04.8(10)   & 6 \\
J1847$-$0402 & B1844$-$04 & 53902 & 18:47:22.8428(16) & $-$04:02:14.26(6)   & 48736 & 18:47:22.8379(18) & $-$04:02:14.19(11)  & 1 \\
J1848$-$0123 & B1845$-$01 & 53868 & 18:48:23.5895(15) & $-$01:23:58.33(5)   & 50022 & 18:48:23.5917(17) & $-$01:23:58.23(7)   & 1 \\
J1901$+$0331 & B1859$+$03 & 54336 & 19:01:31.781(2)   & $+$03:31:05.97(7)   & 50027 & 19:01:31.781(3)   & $+$03:31:05.92(7)   & 1 \\
\\
J1903$+$0135 & B1900$+$01 & 54379 & 19:03:29.9826(11) & $+$01:35:38.37(4)   & 48741 & 19:03:29.981(3)   & $+$01:35:38.33(11)  & 1 \\
J1904$+$0004 & -          & 54377 & 19:04:12.7180(15) & $+$00:04:05.29(4)   & 50866 & 19:04:12.720(7)   & $+$00:04:05.3(3)    & 1 \\
J2022$+$2854 & B2020$+$28 & 53901 & 20:22:37.0685(12) & $+$28:54:22.90(2)   & 49692 & 20:22:37.0671(7)  & $+$28:54:23.104(11) & 1 \\
J2113$+$4644 & B2111$+$46 & 53915 & 21:13:24.328(4)   & $+$46:44:08.82(5)   & 46614 & 21:13:24.307(16)  & $+$46:44:08.70(18)  & 1 \\
J2257$+$5909 & B2255$+$58 & 53908 & 22:57:57.7531(10) & $+$59:09:14.803(8)  & 50085 & 22:57:57.744(4)   & $+$59:09:14.83(3)   & 1 \\
 \noalign{\smallskip}\hline
 \end{tabular}
\end{table*}

Both the frequentist and Bayesian methods assume that the timing noise can be modelled using a power-law spectral model.  This is not true if the pulsar data set contains one or more unmodelled large glitches. PSR J0358+5413 is reported to suffer six glitches\footnote{http://www.atnf.csiro.au/people/pulsar/psrcat/glitchTbl.html \& http://www.jb.man.ac.uk/pulsar/glitches/gTable.html}, four of which happened within our data span.  As the four events are small with $\Delta \nu$/$\nu$ is < = $2e-9$, it is difficult for us to distinguish them from timing noise, although they are detectable with larger telescopes (e. g. Espinoza et al. 2011).  We assumed that these small glitches could be taken as part of the timing noise. Such an assumption was discussed by Coles et al. (2011) and shown, for small glitches, not to significantly bias the resulting estimates.  In order to confirm this we also chose the largest interval (MJD 53219-56645, 2004 Aug 2 -2013 Dec 19) in the data span without a glitch event and repeated the analysis.  With the glitches the position and proper motion estimates are respectively $\rm RA(\rm J2000)= 03^{\rm h}58^{\rm m}53^{\rm s}.7238(2)$, $\rm Dec.(\rm J2000)= +54\degr13'13''.784(3)$ on MJD 54096 and $\mu_{\alpha} = 9.3(5)\,$mas$\,$yr$^{-1}$, $\mu_{\delta} = 8.3(8)\,$mas$\,$yr$^{-1}$. For the restricted data span, they are $\rm RA(\rm J2000)= 03^{\rm h}58^{\rm m}53^{\rm s}.7239(3)$,  $\rm Dec.(\rm J2000)= +54\degr13'13''.782(4)$ on the same epoch\footnote{In order to do the comparison between the two solutions, the epoch of positions measured is fixed on MJD 54096, the integral middle of the whole data span.} and $\mu_{\alpha} = 9.7(7)\,$mas$\,$yr$^{-1}$, $\mu_{\delta} = 8.6(13)\,$mas$\,$yr$^{-1}$.  The glitch events therefore do not significantly bias the position and proper motion estimates.  Accordingly we also ignore the possibility of other small, unpublished glitches in our data set and instead treat such signal as timing noise.

In a few cases (particularly PSR~J1733$-$3716 and J1835$-$1020) the previous position determination was so poor that we could clearly identify an annual sinusoid in our timing residuals when using the catalogue position. The iterative procedure that we have used accounts for such errors and after the iterative procedure we obtain a much improved position estimate.

\subsection{Proper motions}\label{sec:pm}

The proper motions obtained via the two methods are listed in Table~\ref{tab:pm_comp}. Proper motions in right ascension and declination are quoted in mas/yr, i.e., $\mu_{\alpha}$ = $\dot\alpha \cos\delta$ and $\mu_{\delta}$ = $\dot\delta$.  Superscripts `F' and `B' respectively indicate the measurements from the frequentist and the Bayesian methods. The results from both methods are remarkably consistent - both in terms of the parameter values and in terms of their uncertainties. This is not a surprise, the data are the same for the two methods, but does indicate that, even for young pulsars whose residuals are dominated by timing noise, the two methods (Bayesian and frequentist) and noise modelling techniques lead to statistically identical results.

\begin{table*}\scriptsize
\centering
\caption{A comparison of proper motions for 15 pulsars in equatorial coordinates. In column order, the table gives the pulsar name, the proper motions determined by the frequentist and the Bayesian methods, the proper motions derived by comparison of positions given in Table~\ref{tab:pos}, the proper motion determined by the previous timing method, the proper motion measured by the previous interferometric method and its reference (shared with Table~\ref{tab:pos}). Proper motions are quoted in mas/yr, i.e., $\mu_{\alpha}$ = $\dot\alpha \cos\delta$ and $\mu_{\delta}$ = $\dot\delta$. Superscripts `F', `B', `CP', `T' \& `I' seperately denote the frequentist, the Bayesian, the comparison of positions, other timing and the interferometric methods. }
 \label{tab:pm_comp}
 \begin{tabular}{ccccccccccccc}
 \hline\noalign{\smallskip}
PSR J & PSR B & $\mu_{\alpha}^{\rm {F}}$ & $\mu_{\delta}^{\rm {F}}$ & $\mu_{\alpha}^{\rm {B}}$ & $\mu_{\delta}^{\rm {B}}$ & $\mu_{\alpha}^{\rm {CP}}$ & $\mu_{\delta}^{\rm {CP}}$ & $\mu_{\alpha}^{\rm {T}}$ & $\mu_{\delta}^{\rm {T}}$ & $\mu_{\alpha}^{\rm {I}}$ & $\mu_{\delta}^{\rm {I}}$ & Ref.$^{\rm {I}}$ \\
 & & (mas/yr) & (mas/yr) & (mas/yr) & (mas/yr) & (mas/yr) & (mas/yr) & (mas/yr) & (mas/yr) & (mas/yr) & (mas/yr)&   \\
 \hline\noalign{\smallskip}
J0134$-$2937 & -          & 13(2)   &$-$11(3) & 13(2)    & $-$10(3) & 18(2)  & $-$14(3)& -      & -       & -        & -          & - \\
J0358$+$5413 & B0355$+$54 & 9.3(5)  & 8.3(8)  & 10.4(6)  & 7(1)     & 9.2(5) & 8.2(8)  & 8(3)   & 19(6)   & 9.20(18) & 8.17(39)   & 2 \\
J0454$+$5543 & B0450$+$55 & 55(2)   & $-$19(4)& 54(3)    & $-$14(5) & 56(2)  & $-$19(3)& 48(6)  &$-$13(12)& 53.34(6) &$-$17.56(14)& 3 \\
J0953$+$0755 & B0950$+$08 & $-$1(4) & 34(10)  & $-$4(6)  & 25(15)   & $-$2(2)& 30(4)   & $-$3(3)& 26(7)   &$-$2.09(8)  & 29.46(7)   & 7 \\
J1733$-$3716 & B1730$-$37 & 4(9)    & 63(34)  & 6(10)    & 84(37)   & 6(4)  & $-$38(19)& -      & -       & -        & -          & - \\
\\
J1745$-$3040 & B1742$-$30 & 12.5(15)& 30(11)  & 11.9(16) & 50(12)   & 24(6)  & 75(49)  & 6(3)   & 4(26)   & -        & -          & - \\
J1835$-$1020 & -          & 25(5)   & 4(21)   & 24(5)    & 2(21)    & 51(22) & 38(127) & -      & -       & -        & -          & - \\
J1847$-$0402 & B1844$-$04 & $-$1(7) & 8(19)   &$-$0.1(69)& 7(18)    & 5(3)   & 5(9)    & $-$1(5)& $-$9(19)& -        & -          & - \\
J1848$-$0123 & B1845$-$01 & $-$5(6) & 14(16)  & $-$3(6)  & 24(16)   & $-$3(3)& $-$10(8)& 2(6)  & $-$41(19)& -        & -          & - \\
J1901$+$0331 & B1859$+$03 & $-$7(15)& 34(31)  & $-$10(13)& 42(27)   & 0(5)   & 4(8)    & -      & -       & -        & -          & - \\
\\
J1903$+$0135 & B1900$+$01 & 3(7)    &$-$13(14)& 5(12)    & $-$24(28)& 2(4)   & 2(10)   & -      & -       & -        & -          & - \\
J1904$+$0004 & -          & 8(9)    &$-$7(16) & 9(10)    & $-$10(18)& 3(11)  &$-$1(31)&$-$7(21)&$-$83(75) & -        & -          & - \\
J2022$+$2854 & B2020$+$28 & $-$9(4) &$-$19(5) & $-$7(5)  & $-$18(7) & 2(2)   & $-$18(2)& $-$9(3)& $-$15(5)& $-$4.4(5)    & $-$23.6(3)  & 7 \\
J2113$+$4644 & B2111$+$46 & $-$4(13)& 9(14)   & 5(13)    & 4(14)    & 11(8)  & 6(9)    & 9(15)  & $-$1(16)& -        & -          & - \\
J2257$+$5909 & B2255$+$58 & 6(2)    & $-$3(3) & 7(2)     & $-$2(3)  & 7(3)   & $-$3(3) & 28(7)  & $-$8(6) & -        & -          & - \\
 \noalign{\smallskip}\hline
 \end{tabular}
\end{table*}

We also compared the positions in Table~\ref{tab:pos} to derive the proper motions (superscript `CP'). They are all consistent with results obtained by the frequentist and Bayesian methods except for the $\mu_{\delta}$ of three pulsars (PSRs~J1733$-$3716, J1745$-$3040 \& J1835$-$1020) which have poor accuracy on their ealier declination measurements. The remaining part of Table~\ref{tab:pm_comp} lists proper motions and their references from previous publications. The ninth and tenth columns list proper motions available from other timing measurements (superscript `T'). They are all from Hobbs et al. (2004) except for PSRs~J1745$-$3040 and J2022$+$2854 which came from Zou et al. (2005). For four pulsars, i.e., PSRs~J1733$-$4716, J1835$-$1020, J1901$+$0331 and J1903$+$0135, we present the first measurements of their proper motions. The proper motions of PSRs~J0134$-$2937, J1745$-$3040, J1848$-$0123, J1904$+$0004 and J2257$+$5909 are improved and more precise than other timing values. The proper motion of PSR~J0953$+$0755 is comparable to other timing values. That could result from the longer data span of 34.3 yr (Hobbs et al. 2004) for this pulsar. The last three columns present the proper motions from interferometry (superscript `I') and their references. The proper motions measured by the frequentist and the Bayesian methods show the good agreement with the interferometry and other timing results.

Proper motions are generally given in equatorial coordinates. However, timing proper motions are more naturally determined in ecliptic coordinates, especially for pulsars near the ecliptic plane.  For comparison, we also fit ecliptic proper motions for five pulsars that have previously published ecliptic coordinate proper motions, i.e., $\mu_{\lambda}$=$\dot\lambda \cos\beta$ and $\mu_{\beta}$ = $\dot\beta$. These are listed in the third and fourth columns of Table~\ref{tab:pm5}. Measurements from Hobbs et al. (2005) are presented in last two columns (superscript `c'). Our results are generally consistent with the previous measurements.

\begin{table}\scriptsize
\centering
\caption{ The comparison of proper motions for five pulsars which have previously published proper motions in ecliptic coordinates. }
 \label{tab:pm5}
 \begin{tabular}{cccccc}
  \hline\noalign{\smallskip}
PSR J & PSR B & $\mu_{\lambda}^{\rm {F}}$ & $\mu_{\beta}^{\rm {F}}$ & $\mu_{\lambda}^{\rm {c}}$ & $\mu_{\beta}^{\rm {c}}$ \\
      &       & (mas/yr) & (mas/yr) & (mas/yr) & (mas/yr)  \\
  \hline\noalign{\smallskip}
J0134$-$2937 & -          & 6(2)      & $-$15(3)& 8(3)      & $-$21(5)   \\
J1847$-$0402 & B1844$-$04 & $-$0.2(67)& 9(19)   & $-$3(6)   & -          \\
J1848$-$0123 & B1845$-$01 & $-$3(5)   & 16(14)  & $-$0.7(59)& -          \\
J2113$+$4644 & B2111$+$46 & 5(12)     & 11(14)  & 7(12)     & $-$11(13)  \\
J2257$+$5909 & B2255$+$58 & 3(2)      & $-$5(3) & 11(4)     & $-$14(5)   \\
  \noalign{\smallskip}\hline
 \end{tabular}
\end{table}

\subsection{Velocities}

We calculated the pulsar transverse velocities for the pulsars in our sample. The result for seven pulsars in which $\mu_{\rm {tot}}$ is measured with reasonable precision (value/error > 3) is given in Table~\ref{tab:pm_vt}. Followed the pulsar's name, the total proper motion in the equatorial coordinate system was determined by $\mu_{\rm {tot}}$ = $\sqrt{\mu_{\alpha}^{2} + \mu_{\delta}^{2}}$, where $\mu_{\alpha}$ = $\dot\alpha \cos\delta$ and $\mu_{\delta}$ = $\dot\delta$ (using the results obtained with the frequentist method). The best distance estimate (NE2001) for the pulsar from the ATNF pulsar catalogue (Version 1.54) is listed in the fourth column.  The last column contains the transverse velocity $V_{\rm {T}}$ = $4.74\mu_{\rm {tot}}D$, where $D$ is the distance in kpc. Assuming a 20\% error on the distances, we quote uncertainties on the derived velocities.

\begin{table}\scriptsize
\centering
\caption{Velocities for seven pulsars for which $\mu_{\rm {tot}}$ is measured with reasonable
         precision (value/error > 3).}
 \label{tab:pm_vt}
 \begin{tabular}{ccccc}
  \hline\noalign{\smallskip}
PSR J & PSR B & $\mu_{\rm {tot}}$     & Distance & $V_{\rm {T}}$    \\
      &       & (mas/yr)              & (kpc)    & (km$\,$s$^{-1}$) \\
  \hline\noalign{\smallskip}
J0134$-$2937 & -          & 17(3)  & 0.56 & 45(12)   \\
J0358$+$5413 & B0355$+$54 & 12.5(9)& 1.00 & 59(13)   \\
J0454$+$5543 & B0450$+$55 & 58(3)  & 1.18 & 324(67)  \\
J0953$+$0755 & B0950$+$08 & 34(10) & 0.26 & 42(15)   \\
J1745$-$3040 & B1742$-$30 & 33(11) & 0.20 & 31(12)   \\
J1835$-$1020 & -          & 25(8)  & 2.30 & 273(103) \\
J2022$+$2854 & B2020$+$28 & 21(6)  & 2.10 & 209(73)  \\
  \noalign{\smallskip}\hline
  \end{tabular}
\end{table}

\section{Discussion}

\subsection{Spectral index}

 During the process described above, it is necessary to determine the properties of the noise.  This is carried out ``by hand'' in the frequentist method as the user is asked to input parameters for the simple model (equation~\ref{eq:pf}) of the red noise and directly as part of Bayesian automated procedure.  A simple-to-compare measure of the timing noise model is the spectral index of the noise\footnote{Two ways are currently provided to describe the red timing noise in TempoNest, the power-law model presented in van Haasteren (2009) and the model independent method introduced in Lentati (2013).}, $\alpha$.  It is given here as a positive value, which is defined by power spectral density of timing residuals (Hobbs et al. 2010)
\begin{equation}
S(f) \propto f^{-\alpha}
\end{equation}
where $f$ is spectral-frequency. Table~\ref{tab:index} shows spectral indices from the two methods. For most of the 15 pulsars the derived spectral indices are similar, although the frequentist method often gives a somewhat steeper spectral index.  However, as shown by Coles et al. (2011) small errors in the assumed power-law parameters of the noise do not significantly affect the resulting parameters and we note that even though the spectral indices in these two methods are slightly different the resulting proper motions are consistent.

\begin{table}\scriptsize
\centering
\caption{Spectral indices from the frequentist and the Bayesian methods. }
 \label{tab:index}
 \begin{tabular}{cccc}
 \hline\noalign{\smallskip}
PSR J & PSR B & $\alpha^{\rm {F}}$ & $\alpha^{\rm {B}}$   \\
 \hline\noalign{\smallskip}
J0134$-$2937 & -          & 3   & 3(2)   \\
J0358$+$5413 & B0355$+$54 & 8   & 6.2(3) \\
J0454$+$5543 & B0450$+$55 & 6.4 & 5.4(6) \\
J0953$+$0755 & B0950$+$08 & 7   & 5.1(5) \\
J1733$-$3716 & B1730$-$37 & 6.9 & 5.2(8) \\
\\
J1745$-$3040 & B1742$-$30 & 7.4 & 6.4(6) \\
J1835$-$1020 & -          & 7   & 6.0(6) \\
J1847$-$0402 & B1844$-$04 & 6.4 & 5.8(5) \\
J1848$-$0123 & B1845$-$01 & 7.5 & 7.0(6) \\
J1901$+$0331 & B1859$+$03 & 7   & 7.2(8) \\
\\
J1903$+$0135 & B1900$+$01 & 7.2 & 4.0(4) \\
J1904$+$0004 & -          & 6.8 & 4.6(11)\\
J2022$+$2854 & B2020$+$28 & 5.6 & 3.2(2) \\
J2113$+$4644 & B2111$+$46 & 5.2 & 6.1(6) \\
J2257$+$5909 & B2255$+$58 & 8   & 7.0(8) \\
 \noalign{\smallskip}\hline
 \end{tabular}
\end{table}

 In order to understand the differences between the frequentist and Bayesian spectral-index estimates in more detail we have simulated 10 realisations of two pulsars, i.e., PSRs~J0454$+$5543 and J2257$+$5909.  First, an ideal fake data set is formed using the \textsc{formIdeal} plugin to \textsc{tempo2} with the real observed ToAs and the ephemeris assuming a user-specified proper motion in right ascension and declination. High-frequency fluctuations (white noise) in the timing residuals are introduced into the ToAs by running the \textsc{addGaussian} plugin. Low-frequency noise is added using a simple model for the red noise spectrum in equation~\ref{eq:pf} by running the \textsc{addRedNoise} plugin. The spectral index of the simulated red-noise is 6.4 and 8.0 for PSRs~J0454$+$5543 and J2257$+$5909 respectively. Finally, the simulated ToAs were obtained by creating a realisation of the noise (both Gaussian noise and red noise) with the \textsc{createRealisation} plugin. We then processed the simulated data set in exactly the same way as a real one.  For these simulated data sets of the two pulsars we obtained statistically consistent values for the proper motions over a wide range of input parameters.   The spectral indices that were obtained from these simulated data sets are listed in Table~\ref{tab:indexsim}.  The first three columns in the table provide the pulsar names and then whether the results are from the frequentist or Bayesian method.  We then provide the resulting spectral exponent from each of the 10 realisations of the simulated data set.  Note that we obtain an uncertainty for the Bayesian method (listed in parentheses after the parameter value), but do not have an uncertainty on the frequentist determination.  We then list (in columns 14, 15 and 16) the mean and standard deviation of the parameters and the simulated spectral index respectively.  We note that:
\begin{itemize}
\item the uncertainty on the Bayesian estimates does agree with the standard deviation of the parameter values.
\item unsurprisingly the human input leads to more quantised spectral indices being reported, but the values are similar to the simulated value (although see below).
\item the frequentist values are biased high (i.e., the slope is reported to be slightly steeper than the simulated value).  However, as noted by Coles et al. (2011) and also from the results in this paper, slight inaccuracies in the red-noise modelling do not significantly bias the parameter estimates.
\end{itemize}

\begin{table*}\scriptsize
\centering
\caption{Spectral indices from the frequentist (F) and Bayesian (B) methods for 10 realisations of simulated data sets. The final three columns display the mean and standard deviation of the parameters and the simulated spectral index respectively. }
\label{tab:indexsim}
 \begin{tabular}{cccccccccccccccc}
 \hline\noalign{\smallskip}
PSR J & PSR B & Method & 1 & 2 & 3 & 4 & 5 & 6 & 7 & 8 & 9 & 10 & $\bar{\alpha}$ & $\sigma_{\alpha}$ & $\alpha_{Sim}$ \\
 \hline\noalign{\smallskip}
J0454$+$5543 & B0450$+$55 & F & 6.5 & 7.0 & 6.0 & 6.7 & 6.7 & 6.7 & 6.8 & 6.7 & 6.7 & 6.8 & 6.6 & 0.3 & 6.4\\
J0454$+$5543 & B0450$+$55 & B & 5.7(5) & 6.4(4) & 5.6(4) & 4.9(7) & 5.8(4) & 5.7(5) & 5.7(5) & 6.1(4) & 5.6(4) & 5.7(4) & 5.9 & 0.4 & 6.4 \\ \\
J2257$+$5909 & B2255$+$58 & F & 8.5 & 8.7 & 8.2 & 8.5 & 8.5 & 8.6 & 8.4 & 8.6 & 8.6 & 8.6 & 8.5 & 0.1 & 8.0\\
J2257$+$5909 & B2255$+$58 & B & 8.2(7) & 8.2(7) & 8.1(7) & 8.0(8) & 8.1(6) & 8.3(6) & 8.2(7) & 8.2(6) & 8.1(6) & 8.2(7) & 8.2 & 0.1 & 8.0 \\
 \noalign{\smallskip}\hline
\end{tabular}
\end{table*}

\subsection{Comparison of positions}

\begin{figure}
\centering

 \begin{minipage}[b]{0.4\textwidth}
 \includegraphics[angle=-90, width=1\textwidth]{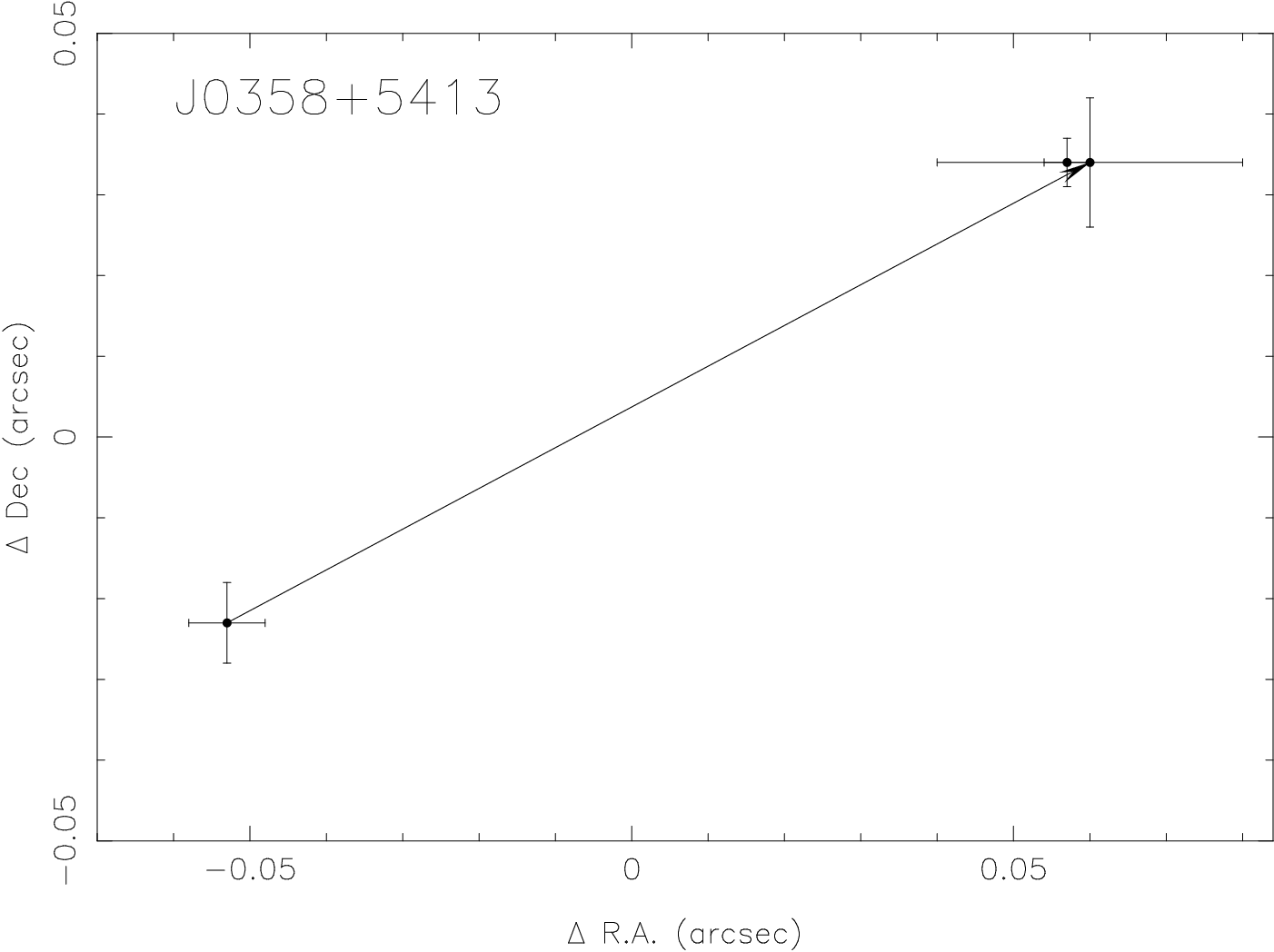}\
 \includegraphics[angle=-90, width=1\textwidth]{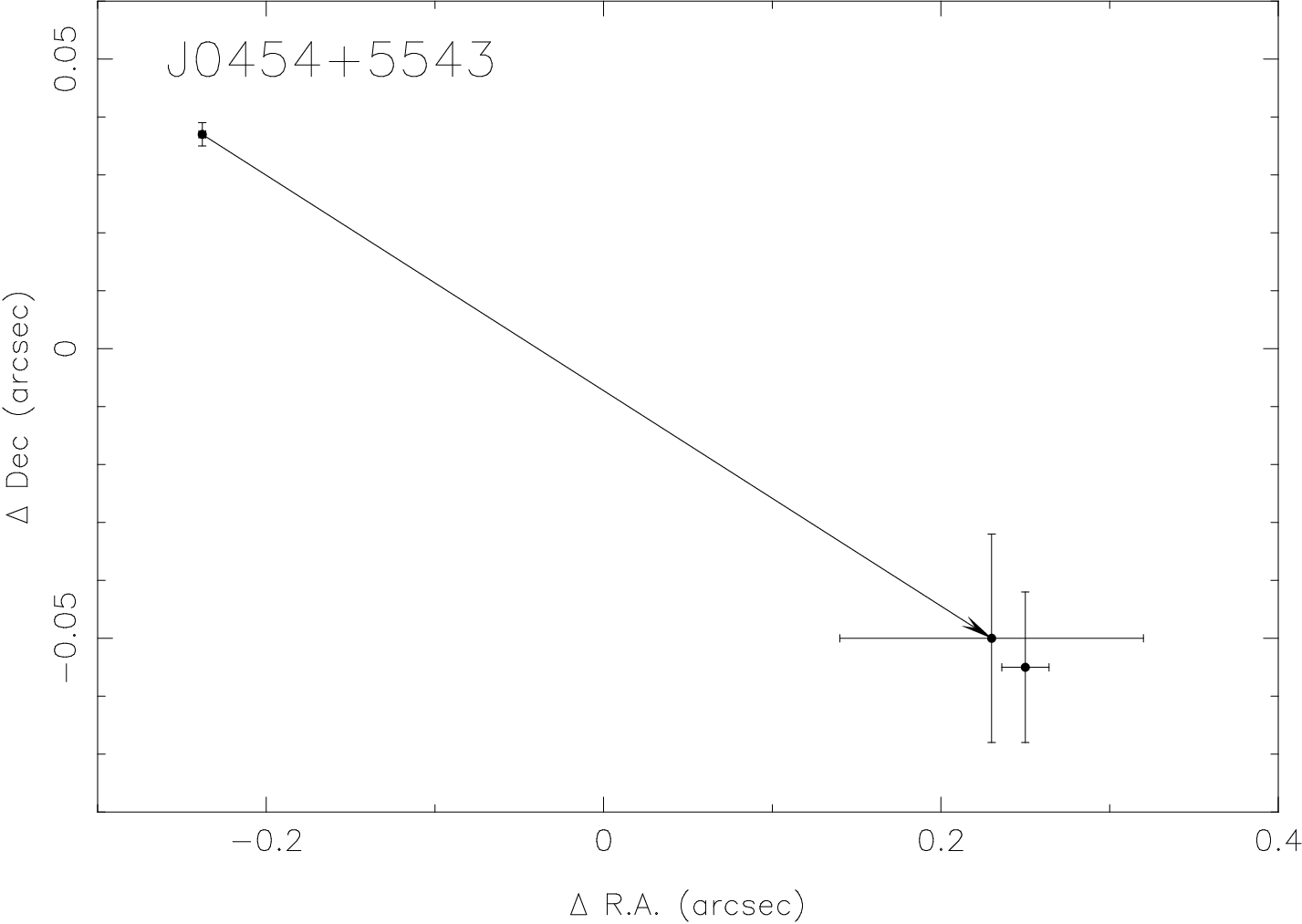}
 \end{minipage}

 \begin{minipage}[b]{0.4\textwidth}
 \includegraphics[angle=-90, width=1\textwidth]{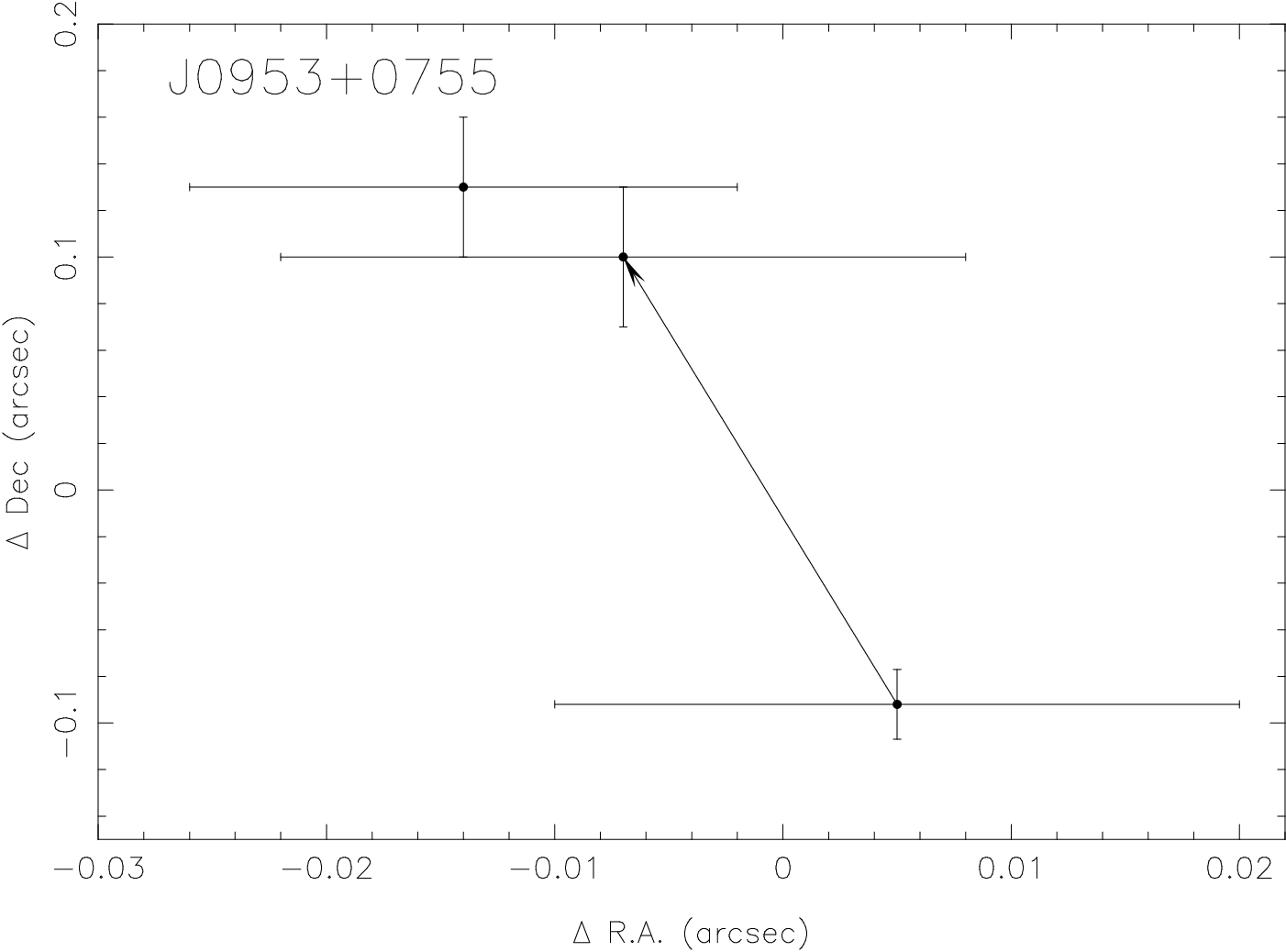}\
 \includegraphics[angle=-90, width=1\textwidth]{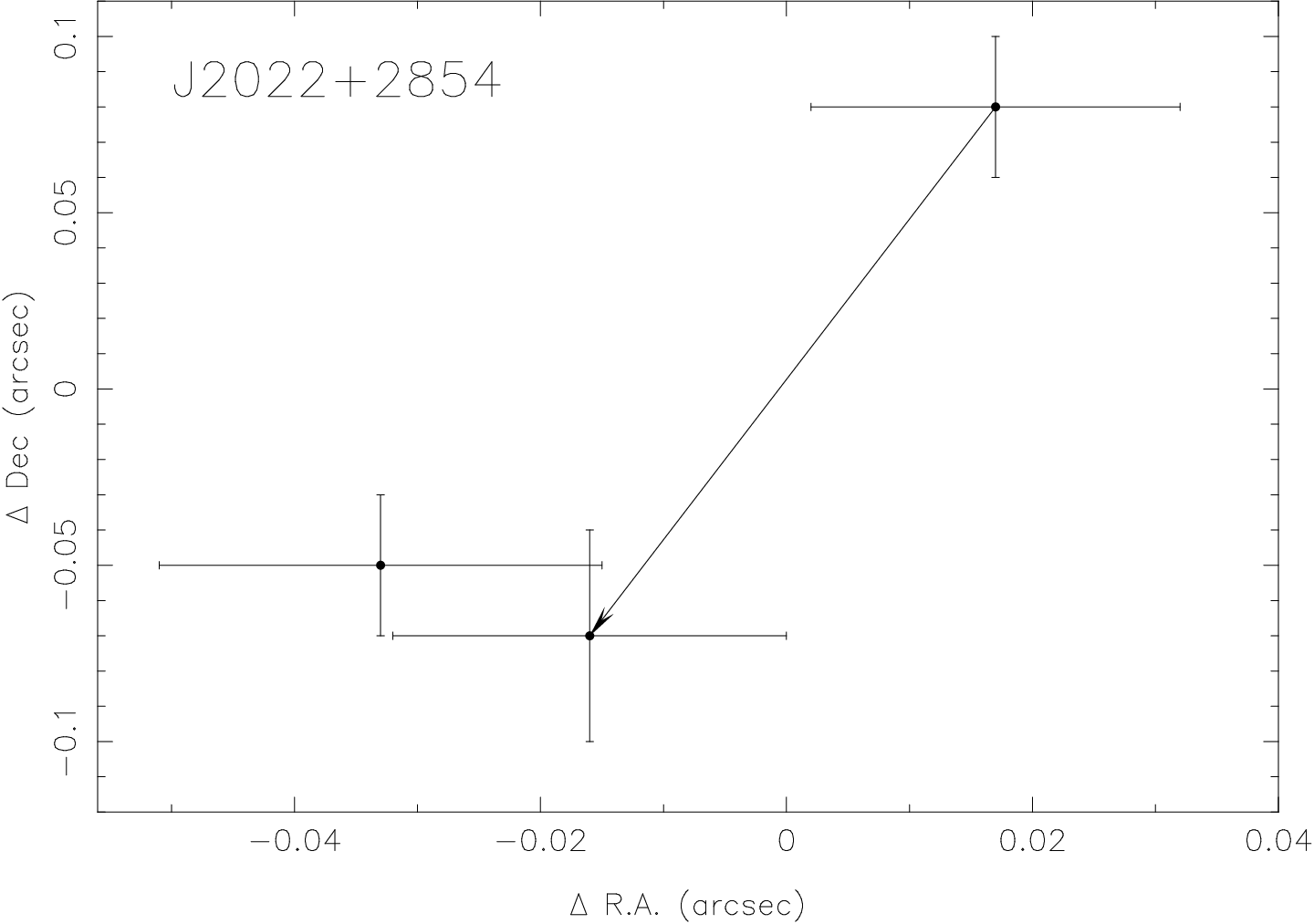}
 \end{minipage}

\caption{Comparisons of positions. For each pulsar, the interferometric position is extrapolated to the epoch of our timing position using the interferometric proper motion. The two interferometric positions are connected by an arrow. Error bars ($\pm 1\sigma$) are plotted for the interferometric positions and the (frequentist) timing positions.}
\label{fig:comp_pos}
\end{figure}

The position measurements for the four pulsars that have published interferometric determinations, i.e., PSR~J0358$+$5413 (Chatterjee et al. 2004), PSR~J0454$+$5543 (Chatterjee et al. 2009), PSR~J0953$+$0755 and PSR~J2022$+$2854 (Brisken et al. 2002), are compared with our (frequentist) results in Fig.~\ref{fig:comp_pos}. Each panel contains the measured interferometric position at the epoch of the measurement joined by an arrow to its expected position at the epoch of our measurement assuming the interferometric proper motion.  We also plot the position obtained by our work.  In all cases our positions are consistent with the extrapolated interferometric positions.

\section{Conclusion}

We have:
\begin{itemize}
\item Obtained proper motions applying the two methods that are not biassed by pulsar timing noise for a sample of 15 pulsars using timing data from the Nanshan observatory. For four pulsars we present the first measurements of their proper motions.
\item Demonstrated that the Coles et al. (2011) and the Lentati et al. (2014) methods can be applied to young pulsar data sets that exhibit large amounts of timing noise with steep spectral exponents.
\item Shown that the two methods give consistent results.
\item Shown that the two methods give results that are consistent with previous very-long-baseline interferometry positions.
\end{itemize}

The two methods give consistent results, but the application procedures for the two methods are very different.  The frequentist method is very hands-on and uses an iterative procedure.  Throughout this procedure the user interacts directly with the data and can identify unexpected features that may be present.   However, it is difficult to make the process reproducible as different users are likely to select different red noise models, use different inputs to the white noise modelling and iterate for a different number of times. In contrast the Bayesian method is quite straightforward - the user sets up an input file and then leaves it to run until the final results are returned. However, the Bayesian procedure requires significant processing power. Therefore, given the same timing models, the frequentist method and the Bayesian method are equivalent in timing analysis, but different in implementation.

Observations of these and other pulsars are on-going at the Nanshan observatory. Over the following years more frequent observations and longer data spans will enable higher accuracy proper motions and position determinations. In the future, the Five hundred meter Aperture Spherical radio Telescope (FAST) and the QiTai radio Telescope (QTT) pulsar surveys will discover more pulsars and timing programs on these telescopes will be able to determine the astrometric parameters with much higher precision than those available to us with our existing telescopes.

\section*{Acknowledgements}

We would like to thank the reviewer very much for helpful comments. We thank the staff who helped with the observations and the engineers responsible for maintaining the receiver and telescope at XAO. We thank Dr. Meng Yu for his helpful discussion on data analysis. This work was supported by National Basic Research Program of China (973 Program 2015CB857100 and 2012CB821801), the Strategic Priority Research Programme (B) of the Chinese Academy of Sciences (No. XDB09000000), National Natural Science Foundation of China (No. 11463005, 11403086 and U1431107) and West Light Foundation of CAS (No. XBBS201322). G. Hobbs was supported by the Australian Research Council (ARC) Future Fellowship. L. Lentati was supported by a Junior Research Fellowship at Trinity Hall College, Cambridge University.

%%%%%%%%%%%%%%%%%%%%%%%%%%%%%%%%%%%%%%%%%%%%%%%%%%



%\bibliographystyle{mnras}


%%%%%%%%%%%%%%%%%%%%%%%%%%%%%%%%%%%%%%%%%%%%%%%%%%

%%%%%%%%%%%%%%%%% APPENDICES %%%%%%%%%%%%%%%%%%%%%

%%%%%%%%%%%%%%%%%%%%%%%%%%%%%%%%%%%%%%%%%%%%%%%%%%

% Don't change these lines
\bsp	% typesetting comment
\label{lastpage}
\end{document}